\documentclass[aps,prd,twocolumn,showpacs,floatfix,nofootinbib,superscriptaddress]{revtex4-2}
\usepackage{dcolumn}

\usepackage{amsmath,amsfonts,extarrows,amssymb,mathrsfs,times,bm}
\usepackage{extarrows}
\usepackage{graphicx}
\usepackage{float}
\usepackage{graphicx,epstopdf}
\usepackage{dcolumn}
\usepackage{exscale}
\usepackage{relsize}
\usepackage{mathtools}
\usepackage{mhchem}
\usepackage{changes}
\usepackage[colorlinks=true,breaklinks=true,linkcolor=blue,citecolor=blue,urlcolor=blue]{hyperref}
\newcommand{\nn}{\nonumber}

\newcommand{\im}{\mathrm{i}}
\newcommand{\dut}[3]{#1_{#2}^{\phantom{#2}#3}}

\begin{document}

\title{Kerr black hole surrounded by a cloud of strings and its weak gravitational lensing in Rastall gravity}
\author{Zonghai Li}
\email{lzh@my.swjtu.edu.cn}
\affiliation{School of Physical Science and Technology, Southwest Jiaotong University, Chengdu 610031, China}
\affiliation{MOE Key Laboratory of Artificial Micro- and Nano-structures, School of Physics and Technology, Wuhan University, Wuhan 430072, China}
\date{\today}
\author{Tao Zhou}
\email{taozhou@swjtu.edu.cn}
\affiliation{School of Physical Science and Technology, Southwest Jiaotong University, Chengdu 610031, China}

\begin{abstract}
In this paper, an exact solution of Kerr black hole surrounded by a cloud of strings in Rastall gravity is obtained by Newman-Janis algorithm without complexification, and the influence of the string parameter $a_0$ on the black hole thermodynamics is studied. Furthermore, according to this black hole solution, we investigate the effect of a cloud of strings on gravitational deflection of massive particles. For a receiver and a source at infinite distance from the lens object, we use an osculating Riemannian manifold method. While the distance is finite, we apply the generalized Jacobi metric method. For the case between the two situations, the Jacobi-Maupertuis Randers-Finsler metric and Gauss-Bonnet theorem are employed. We find that the string parameter $a_0$ has obviously modification on the gravitational deflection of massive particles. Our result is reduced to the deflection angle of light by Kerr black hole surrounded by a cloud of strings in general relativity.
\end{abstract}

\maketitle

\section{Introduction}
 Einstein's general relativity (GR) is a beautiful theory of gravity. It has undergone a series of tests and became the basis of modern astrophysics and cosmology. For GR, there is an important assumption about conservation laws due to equivalence principle: the covariant divergence of the energy-momentum tensor is vanishing. Although this conservation law has been tested specifically in a weak gravitational field case, it may not hold in strong gravitational field. Assuming that the covariant divergence of the energy-momentum tensor is non-zero, Rastall proposed an extended theories of gravity~\cite{Rastall1972}. This non-conservation gravity theory may provide a model to explain the problems of dark energy and dark matter. Moreover, Rastall gravity seems to agree with the observational data of the universe age, Hubble parameter and helium nucleosynthesis~\cite{Al-Rawaf1996,Al-Rawaf2005}. Hence, it is still an important theory of gravity. Recent interests of various areas within the context of the Rastall gravity or the generalized Rastall theory can be found in the literatures~\cite{Moradpour2019,Moraes2019,Bamba12018,Yu2019,Darabi12018,Visser2018,Darabi2018,Ziaie2019}.

On the other hand, black hole (BH) is one of the most interesting predictions of GR. In 2015, gravitational waves from binary BH mergers were directly detected by LIGO and Virgo collaborations~\cite{Abbott2016}, which is a crucial proof of the existence of BHs. Furthermore, the first image of the BH was  observed in 2019 by the Event Horizon Telescope collaboration~\cite{Akiyama2019L1,Akiyama2019L4}. For Rastall gravity model, some exact solutions have been discovered. For example, Heydarzade \textit{et al.} found a static charged BH solution~\cite{Heydarzade2017}. Kumar \textit{et al.} obtained the rotating BH solution~\cite{Kumar:2018Kerr,Kumar:2017Kerrshadow}. Moradpour \textit{et al.} obtained the traversable asymptotically flat wormholes~\cite{Moradpour2017}. Furthermore, the BHs surrounded by different matter were proposed in Rastall gravity. Heydarzade and Darabi obtained a class of solutions of Kiselev-like BHs surrounded by perfect fluid~\cite{Heydarzade2017:BHs}, where Kiselev's original work can be found in~\cite{Kiselev2003}. Xu \textit{et al.} found the solution of Kerr-Newman-AdS BH surrounded by perfect fluid matter~\cite{Xu2018}.

The string theory believes that the basic unit of nature is extended one-dimensional strings, instead of the point-like particles of particle physics. The gravitational field induced by a collection of strings is worth studying, because we can test the basic theory via investigating the gravitational effect. A cloud of strings as the source of gravitational field was first considered by Latelier, within the context of GR~\cite{Letelier1979,Letelier1981,Letelier1983}. He found an exact solution of Schwarzschild BH surrounded by a cloud of strings. Later, the rotating BHs with cloud of strings were investigated~\cite{Barbosa,ToledoBezerra}. Nowadays, the study of a cloud of strings was extended to modified theories of gravity such as Lovelock gravity and $f(R)$ gravity~\cite{Herscovich2010,Toledo2019,Morais2018}. Quite recently, Cai and Miao found an exact solution of Schwarzschild black hole surrounded by a cloud of strings in Rastall gravity and investigated its quasinormal modes and spectra~\cite{Cai2019}. Actually, rotating BH solutions are more general and can provide more applications in astrophysics and cosmology. Thus, in this paper, we try to give an exact solution of Kerr BH surrounded by a cloud of strings in Rastall gravity through Newman-Janis algorithm (NJA) without complexification.

The gravitational lensing is one of the most powerful tools in astrophysics and cosmology, and due to its importance in testing fundamental theory of gravity~\cite{DED1920,Will2015}, measuring the mass of galaxies and clusters~\cite{Hoekstra2013,Brouwer2018,Bellagamba2019}, detecting dark matter and dark energy~\cite{Vanderveld2012,cao2012,zhanghe2017,Huterer2018,SC2019,Andrade2019}, and so on, in the present work, we will mainly investigate gravitational lensing for the Kerr BH surrounded by a cloud of strings.

Typically, in most studies involving the gravitational deflection of light, the standard geodesics method was mainly considered~\cite{Weinberg,Edery1998,Virbhadra1998,Virbhadra2000,Virbhadra2002,Virbhadra2008,Virbhadra2009,Bodenner2003,Nakajima2012,Cao&Xie:light2018,Wang2019}. Recently, Gibbons and Werner introduced an interesting geometric method in Ref.~\cite{GW2008}. They applied the Gauss-Bonnet (GB) theorem to study the weak gravitational deflection of light in static and spherically symmetric (SSS) spacetime such as Schwarzschild spacetime. In their method, the deflection angle of light can be calculated by integrating Gaussian curvature of corresponding optical metric. Later, Bloomer~\cite{Bloomer2011} attempted to extend this geometrical approach to the stationary and axially symmetric (SAS) spacetime, where the corresponding optical geometry is defined by a Randers-Finsler metric. Until 2012, Werner~\cite{Werner2012} completely solved this problem by building an osculating Riemannian manifold of the Randers-Finsler manifold, with Naz{\i}m's method~\cite{Nazim1936}. Importantly, Gibbons-Werner (GW) method shows that the gravitational lensing can be viewed as a global effect. In addition, this method only involves spatial geometry (optical geometry), so it is conducive to the implementation of physical lens models~\cite{Werner2012}. At last, it contributes to the physical applications of Finsler geometry and geometric dynamics (Jacobi metric).

With the GW method, the light deflection in different spacetimes have been widely studied by some authors. For example, Jusufi~\textit{et al.} studied the deflection of light by BHs, wormholes and other lens objects ~\cite{Jusufi:wormhole,Jusufi:string17,Jusufi&Ali:Teo,Jusufi:string,Jusufi&Ali:string,Jusufi:RB}. \"{O}vg\"{u}n~\textit{et al.} studied light deflection in asymptotically non-flat spacetime such as the Schwarzschild-like spacetime in bumblebee gravity model~\cite{Ali:wormhole,Ali:strings,Ali:BML}. Javed \textit{et al.} studied the effect of different matter fields on weak gravitational deflection of light~\cite{Javed1,Javed2,Javed3,Javed4}. More works can also be found in Refs.~\cite{Sakalli2017,Goulart2018,Leon2019,zhu2019}. In addition, massive particle is an important class of element particles in the universe, and the analysis of the signatures of the gravitational lensing will be useful to understand the properties of these particles. Due to the importance of this problem, various works~\cite{AR2002,AP2004,Bhadra2007,Yu2014,He2016,He2017a,He2017b,Jia2016,Jia2019,Jia&Liu:2019} have been carried out to study gravitational deflection of massive particles by different lens objects in differential gravity models. With GW method, Crisnejo~\textit{et al.} studied not only the weak gravitational deflection of light in a plasma medium, but also the deflection of massive particles, in SSS/SAS spacetimes~\cite{CG2018,CGV2019,CGJ2019}. Jusufi studied the deflection angle of massive particles in Kerr spacetime and Teo spacetimes, as well as the deflection angle of charged massive particles in Kerr-Newman spacetime~\cite{Jusufi:mp,Jusufi:cmp}. Moreover, with Jacobi metric method, Li~\textit{et al.} studied the gravitational deflection of massive particles in wormhole spacetimes~\cite{LHZ2019}.

Inspired by the GW method, on the other hand, the finite-distance gravitational deflection of light have been studied by some authors, where the receiver and the source are assumed to be at finite distance from a lens. In 2016, Ishihara \textit{et al.}~\cite{ISOA2016,IOA2017} used the GB theorem to study the finite-distance deflection of light in SSS spacetime. Later, Ono \textit{et~al.}~\cite{OIA2017,OIA2018,OIA2019} proposed a generalized optical metric method to extend the work to SAS spacetime. It is worthwhile to mention that the generalized optical metric method can also be used to calculate the infinite-distance deflection angle~\cite{Ali:OIA1,Ali:OIA2,LZ2019,Kumar2019}. A review on finite-distance deflection of light can be found in Ref.~\cite{OA2019}. Furthermore, Arakida~\cite{Arakida2018} studied the finite-distance deflection of light in Schwarzschild-dS spacetime. In addition, Crisnej~\textit{et~al.} investigated the finite-distance deflection of light in SSS gravitational field with a plasma medium~\cite{CG2019}. Haroon \textit{et al.} studied the finite-distance deflection of light by a rotating BH in perfect fluid dark matter with a cosmological constant~\cite{Haroon2019}. Quite recently, by using Jacobi-Maupertuis-Randers-Finsler (JMRF) metric and GB theorem, Li~\textit{et al.} studied the finite-distance gravitational deflection of massive particles both in asymptotically flat spacetime~\cite{LJ2019} and asymptotically non-flat spacetime~\cite{LA2019}.

To test basic gravitational theory, or in other cosmological applications, the lens effect of a cloud string clouds is very interesting and worth studying. Jusufi~\textit{et al.}~\cite{Jusufi:monopole} has studied the effect of cloud of strings on deflection angle of light by Kerr in GR, using GB theorem. In this paper, we will extend this result via studying the deflection of massive particle by Kerr BH surrounded by a cloud of strings in Rastall gravity, using GB theorem. The aim of the present work is twofold. On the one hand, we will extend the static BH solution to the rotating case solution by NJA without complexification. On the other hand, we will study the effects of a cloud of string on both thermodynamic properties and weak gravitational deflection of massive particles. In particular, we study the deflections for infinite-distance case with Werner's method, and for finite-distance case with the generalized Jacobi metric method.

This paper is organized as follows. In Sec.~\ref{KerrRastll}, we first review the Schwarzschild BH surrounded by a cloud of strings in Rastall gravity theory. Then, by using the NJA without complexification, we derive an exact solution which represents a rotating Kerr BH surrounded by a cloud of strings in Rastall gravity. Thereafter, the thermodynamic properties of this rotating BH are discussed. Furthermore, according to this solution, in Sec.~\ref{InGBDEF} the GB theorem and osculating Riemannian manifold method are used to study the effects of a cloud of strings on gravitational deflection angle of massive particles for a receiver and a source at infinite distance from the lens. In Sec.~\ref{GBDEF}, we use GB theorem and the generalized Jacobi metric method to study the case for a receiver and source at a finite distance from the lens. Finally, we end our paper with a short conclusion in Sec.~\ref{conclusion}. For simplicity, we set $G = c =\hbar=k_B= 1$ in this paper.

\section{Kerr BH surrounded by a cloud of strings in Rastall gravity}\label{KerrRastll}

\subsection{Schwarzschild BH surrounded by a cloud of strings}
\label{Schwarzschild}
Let us begin with the field equation of Rastall gravity model~\cite{Rastall1972}
\begin{eqnarray}
\label{fieldequation}
&&H_{\mu\nu}=G_{\mu\nu}+\beta g_{\mu\nu}\mathcal{R}=\kappa \mathcal{T}_{\mu\nu},\nn\\
&&\mathcal{T}^{\mu\nu}_{~~~~;\mu}=\lambda \mathcal{R}^{,\nu},
\end{eqnarray}
where $\beta\equiv\kappa\lambda$ with the constant $\lambda$ being Rastall parameter and $\kappa$ being the Rastall gravitational coupling constant. It is required that $\beta\neq1/6$ and $\beta\neq1/4$~\cite{MoradpourSalakoA}. Obviously, Eq.~\eqref{fieldequation} is reduced to the field equation and the conservation of the energy-momentum tensor within the context of the general relativity when $\lambda=0$ and $\kappa=8\pi G$.

The two-dimensional world sheet $\Pi$ of a string is described by $x^\alpha=x^\alpha(X^a)$, where $X^0$ is timelike parameter and $X^1$ is spacelike parameter. The action of a string evolving in the spacetime is~\cite{Letelier1979}
\small\begin{eqnarray}
\label{stringaction}
 I_{S}=\mu\int_{\Pi} \sqrt{-\chi} d X^{0} d X^{1}=\mu\int_{\Pi} \sqrt{-\frac{1}{2}\Pi_{\alpha\beta}\Pi^{\alpha\beta}} d X^{0} d X^{1},
\end{eqnarray}
where $\mu$ is a positive dimensionless constant related to the tension of the string, $\chi$ is the determinant of the induced metric
\begin{eqnarray}
\chi_{ab}=g_{\alpha\beta}\frac{\partial x^\alpha}{\partial X^a}\frac{\partial x^\beta}{\partial X^b},
\end{eqnarray}
and $\Pi_{\alpha\beta}$ is the string bi-vector defined as
\begin{eqnarray}
\Pi^{\alpha\beta}=\epsilon^{ab}\frac{\partial x^\alpha}{\partial X^a}\frac{\partial x^\beta}{\partial X^b}
\end{eqnarray}
with $\epsilon^{ab}$ the Levi-Civita symbol satisfying $\epsilon^{01}=-\epsilon^{10}=1$. The energy-momentum tensor of the string can be obtained from the action in Eq.~\eqref{stringaction} as follows
\begin{eqnarray}
T^{\alpha\beta}=\mu \frac{\Pi^{\alpha\gamma}\dut{\Pi}{\gamma}{\beta}}{\sqrt{-\chi}}.
\end{eqnarray}

For a collection of strings, the energy-momentum tensor is $T_\mathrm{cloud}^{\alpha\beta}=\rho_sT^{\alpha\beta}$, where $\rho_s$ is the number density. Considering the spherically symmetric distributions of string cloud, Cai and Miao recently obtained a BH solution to the field equation~\eqref{fieldequation}, as follows~\cite{Cai2019}
\small\begin{eqnarray}
\label{Schwarzschild}
 &&ds^2=-f(r)dt^2+\frac{dr^2}{g(r)}+h(r)\left(d\theta^2+\sin^2\theta d\phi^2\right),
\end{eqnarray}
where
\begin{eqnarray}
&&f(r)=g(r)=1-\frac{2M}{r}+\frac{4a_0 \left(\beta-\frac{1}{2}\right)^2}{\left(8\beta^2+2\beta-1\right)r^{\frac{4\beta}{2\beta-1}}}, \nn\\
&& h(r)=r^2.\nn
\end{eqnarray}
Here, $M$ is the mass of the BH and $a_0$ is string parameter linked to the energy density of the cloud of strings. It is obvious that $\beta\neq \pm 1/2, 1/4$. It is worthwhile to mention that the solution of the Schwarzschild BH surrounded by a cloud of strings in Rastall gravity in Eq.~\eqref{Schwarzschild} can be transformed into the solution of the Schwarzschild BH surrounded by quintessence in GR~\cite{Kiselev2003} by the following transformation~\cite{Cai2019}:
\begin{eqnarray}
\label{transformation}
&&\beta=\frac{3 \omega_{q}+1}{6 \omega_{q}-2}, \nn\\
&&a_0=-\frac{9 \tilde{c}}{2}\left(\omega_{q}+\omega_{q}^{2}\right),
\end{eqnarray}
where $\omega_{q}$ is the quintessential state parameter, and $\tilde{c}$ is an integral constant associated with the energy density of quintessence.

\subsection{Newman-Janis algorithm without complexification and Kerr BH surrounded by a cloud of strings}
\label{Newman}
The NJA is a useful technique to obtain rotating spacetime from a nonrotating spacetime~\cite{NJ01,NJ02,Ghosh2016,Shaikh2019}. For example, one can apply it to construct the Kerr (Kerr-Newman) solution from Schwarzschild (Reissner-Nordstr\"{o}m) solution. However, due to the complexification of the radial coordinate $r$ in the original NJA, the solution obtained by the method may be invalid in the Boyer-Lindquist coordinate system~\cite{Azreg2011}. Azreg-A\"{\i}nou introduced a modified NJA that can drop the complexification~\cite{Azreg20141,Azreg20142}. Azreg-A\"{\i}nou's method of NJA without complexification has now been widely used in different gravitational theories including Rastall model~\cite{Xu2018,Xu2017,Chen2019,Jusufi:NJ2019}. In particular, it is shown that the rotating black hole surrounded by a cloud of strings in GR can be obtained using the modified NJA~\cite{ToledoBezerra}. In this subsection, the NJA without complexification will be employed to build the rotating BH solution from the static one given in Eq.~\eqref{Schwarzschild}.

Via the coordinate transformation
\begin{eqnarray}
&&du=dt-\frac{dr}{\sqrt{f(r)g(r)}},
\end{eqnarray}
the metric in Eq.~\eqref{Schwarzschild} becomes
\begin{eqnarray}
 ds^2=-f(r)du^2-2\sqrt{\frac{f(r)}{g(r)}}du dr+h(r)\left(d\theta^2+\sin^2\theta d\phi^2\right).
\end{eqnarray}
Using the null tetrad, the inverse components of the metric can be written as
\begin{eqnarray}
\label{nullmetric}
 &&g^{\mu\nu}=-l^\mu n^\nu-l^\nu n^\mu+m^\mu \bar{m}^\nu+m^\nu \bar{m}^\mu,
\end{eqnarray}
or in the matrix form
\[g^{\mu\nu}={\begin{pmatrix}
 0&-1&0&0\\
 -1&0&0&0\\
 0&0&0&1\\
 0&0&1&0
\end{pmatrix}},\]
with $\bar{m}^{\mu}$ the complex conjugate of $m^{\mu}$, and one can chose the tetrad vectors as follows
\begin{eqnarray}
&& l^{\mu}=\left(0,1,0,0\right)~,\nn\\
&& n^{\mu}=\left(\sqrt{\frac{g(r)}{f(r)}},-\frac{g(r)}{2},0,0\right)~,\nn\\
&& m^{\mu}=\frac{1}{\sqrt{2h(r)}}\left(0,0,1,\frac{\im}{\sin\theta}\right).
\end{eqnarray}

Now a complex coordinate transformation can be performed on the $(u,r)$ plane
\begin{eqnarray}
&& u\rightarrow u-\im a\cos\theta,\nn\\
&& r\rightarrow r+\im a\cos\theta,
\end{eqnarray}
where $a$ is the spin parameter. The new coordinates $(u,r,\theta,\phi)$ is called Eddington-Finkelstein coordinates. The transformation leads to $f(r)\rightarrow F(r,\theta,a)$, $g(r)\rightarrow G(r,\theta,a)$, and $h(r)\rightarrow\Sigma(r,\theta,a)$. In Edding-Finkelstein coordinates, the null tetrad becomes
\begin{eqnarray}
&& l^{\mu}=\left(0,1,0,0\right),\nn\\
&& n^{\mu}=\left(\sqrt{\frac{G}{F}},-\frac{G}{2},0,0\right),\nn\\
&& m^{\mu}=\frac{1}{\sqrt{2\Sigma}}\left(\im a\sin\theta,-\im a\sin\theta,1,\frac{\im}{\sin\theta}\right).
\end{eqnarray}
Then, according to Eq.~\eqref{nullmetric}, the metric can be obtained and the result reads
\begin{eqnarray}
\label{EFmetric}
ds^2&=&-Fdu^2-2\sqrt{\frac{F}{G}}du dr+2a\left(F-\sqrt{\frac{F}{G}}\right)\sin^2\theta du d\phi\nn\\
&&+\Sigma d\theta^2+2a\sin^2\theta\sqrt{\frac{F}{G}}dr d\phi\nn\\
&&+\sin^2\theta\left[\Sigma+a^2\left(2\sqrt{\frac{F}{G}}-F\right)\sin^2\theta\right]d\phi^2~.
\end{eqnarray}
Next, one can change the Edding-Finkelstein coordinates $(u,r,\theta,\phi)$ to Boyer-Lindquist coordinates $(t,r,\theta,\phi)$ via~\cite{Azreg20141}
\begin{eqnarray}
&&du\rightarrow dt+\xi(r)dr\nn,\\
&&d\phi\rightarrow d\phi+\zeta(r)dr,
\end{eqnarray}
where
\begin{eqnarray}
&&\xi(r)=-\frac{\varsigma(r)+a^2}{g(r)h(r)+a^2},\nn\\
&&\zeta(r)=-\frac{a}{g(r)h(r)+a^2},\nn\\
&&\varsigma(r)=\sqrt{\frac{g(r)}{f(r)}}h(r).\nn\\
\end{eqnarray}
Moreover, one can choose
\begin{eqnarray}
&&F(r,\theta)=\frac{\left(g(r)h(r)+a^2\cos^2\theta\right)\Sigma}{\left(\varsigma(r)+a^2\cos^2\theta\right)^2},\nn\\
&&G(r,\theta)=\frac{g(r)h(r)+a^2\cos^2\theta}{\Sigma}.\nn\\
\end{eqnarray}
For our case, $f(r)=g(r)$ and $h(r)=r^2$, one can choose $\Sigma=r^2+a^2\cos^2\theta$~\cite{Azreg20141}. Thus, one has
\begin{eqnarray}
F(r,\theta)=G(r,\theta)=\frac{r^2g(r)+a^2\cos^2\theta}{\Sigma}.
\end{eqnarray}

Finally, the metric of Kerr BH surrounded by a cloud of strings in the Boyer-Lindquist coordinates reads
\begin{eqnarray}
\label{Kerrmetric}
ds^2&=&-\left(1-\frac{A}{\Sigma}\right)dt^2+\frac{\Sigma}{\Delta}dr^2\nn\\
&&+\Sigma d\theta^2-\frac{2aA\sin^2\theta}{\Sigma}dtd\phi~\nn\\
&&+\sin^2\theta\left(r^2+a^2+\frac{a^2A\sin^2\theta}{\Sigma}\right)d\phi^2,
\end{eqnarray}
where
\begin{eqnarray}
&& A=2Mr-4a_0\lambda_1r^{2-\lambda_2},\nn\\
&&\Delta=r^2-2Mr+a^2+4a_0\lambda_1r^{2-\lambda_2},\nn\\
&&\lambda_1=\frac{\left(\beta-\frac{1}{2}\right)^2}{8\beta^2+2\beta-1},\nn\\
&&\lambda_2=\frac{4\beta}{2\beta-1}.\nn
\end{eqnarray}

One can obtain the nonvanishing components of the Rastall tensor $H_{\mu\nu}$ associated with metric in Eq.~\eqref{Kerrmetric} by using the Mathematica package RGTC, as follows
\begin{eqnarray}
H_{00}&=&\frac{2\left({a}^{2} {r}^{2}+{r}^{4}-2 {r}^{3} \rho-{a}^{4} \cos^2 \theta \sin^2 \theta\right) \rho^{\prime}}{\Sigma^{3}}\nn\\
&&-\frac{{r}\left({a}^{2}+{r}^{2}-2 {r} \rho\right) \rho^{\prime \prime}}{\Sigma^{2}},\nn\\
H_{11}&=&-\frac{2 {r}^{2}}{\Sigma} \frac{\rho^{\prime}}{\Delta}+\frac{{r} \rho^{\prime \prime}}{\Delta},\nn\\
H_{22}&=&-\frac{2 {a}^{2} \cos^2\theta}{\Sigma} \rho^{\prime},\nn\\
H_{33}&=&-\frac{2 a^{2} \sin^2 \theta\left[\left(a^{2}+r^{2}\right)^{2} \cos^2\theta-r^{2}\left(a^{2}+r^{2}-2 r \rho\right) \sin^2\theta\right]\rho^{\prime}}{\Sigma^{3}}\nn\\
&&-\frac{r a^{2}\left(a^{2}+r^{2}-2 r \rho\right) \sin^4 \theta \rho^{\prime \prime}}{\Sigma^{2}},\nn\\
H_{03}&=&\frac{2a\left[\left(a^{2}+r^{2}\right)\left(a^{2} \cos^2 \theta-r^{2}\right)+2 r^{3} p\right] \sin^2 \theta \rho^{\prime}}{\Sigma^{3}} \nn\\
&& +\frac{ar\left(a^{2}+r^{2}-2 r \rho\right) \sin^2\theta \rho^{\prime \prime}}{\Sigma^{2}},\nn
\end{eqnarray}
where $\rho=M-2a_0\lambda_1r^{2-\lambda_2}$ and ``$\prime$'' denotes the derivative with respect to $r$. To calculate the energy-momentum tensor, we use the following tetrad
\begin{eqnarray}
e_0^{\mu}&=&\frac{1}{\sqrt{\Sigma \Delta}}\left(r^2+a^2,0,0,a\right),\nn\\
e_1^{\mu}&=&\sqrt{\frac{\Delta}{\Sigma}}\left(0,1,0,0\right),\nn\\
e_2^{\mu}&=&\frac{1}{\sqrt{\Sigma }}\left(0,0,1,0\right),\nn\\
e_3^{\mu}&=&-\frac{1}{\sqrt{\Sigma}\sin\theta}\left(a\sin^2\theta,0,0,1\right).
\end{eqnarray}
With this tetrad, the energy-momentum tensor  can be written as $\mathcal{T}_{\mu \nu}=(\mathcal{E},\mathcal{P}_1,\mathcal{P}_2,\mathcal{P}_3)$, where
\begin{eqnarray}
&&\mathcal{E}=\frac{1}{\kappa}e_0^{\mu}e_0^{\nu}H_{\mu\nu},~~~~\mathcal{P}_1=\frac{1}{\kappa}e_1^{\mu}e_1^{\nu}H_{\mu\nu},\nn\\
&&\mathcal{P}_2=\frac{1}{\kappa}e_2^{\mu}e_2^{\nu}H_{\mu\nu},~~~~\mathcal{P}_3=\frac{1}{\kappa}e_3^{\mu}e_3^{\nu}H_{\mu\nu}.\nn
\end{eqnarray}
Then, the results read
\begin{eqnarray}
\label{energy-momentum}
&&\mathcal{E}=-\mathcal{P}_1=\frac{r\left(2 r \rho^{\prime}-\Sigma \rho^{\prime \prime}\right)}{\kappa \Sigma^{2}},\nn\\
&&\mathcal{P}_2=\mathcal{P}_3=-\mathcal{P}_1-\frac{2 \rho^{\prime}-r \rho^{\prime \prime}}{\kappa \Sigma}.
\end{eqnarray}
Thus, the solution given by Eq.~\eqref{Kerrmetric} indeed describes a rotating BH surrounded by a cloud of strings with energy-momentum tensor Eq~\eqref{energy-momentum}, within the context of the Rastall gravity. The solution in Eq.~\eqref{Kerrmetric} can return to the various known BH solutions in suitable limits. The Kerr BH surrounded by a cloud of strings in GR can be derived when Rastall parameter $\lambda=0$~\cite{ToledoBezerra}. When $a_0=0$, the energy-momentum tensor Eq~\eqref{energy-momentum} vanishes, and the Kerr solution in GR can be found~\cite{BL1967}. Meanwhile, it returns to the static solution in Eq.~\eqref{Schwarzschild} if $a=0$. As well, using the transformation in Eq.~\eqref{transformation}, the solution of the Kerr BH surrounded by a cloud of strings in Rastall gravity in Eq.~\eqref{Kerrmetric} can be transformed into the solution of the Kerr BH surrounded by quintessence in GR~\cite{ToshmatovEPJC2017}.

In the following, we will study the thermodynamics and weak gravitational deflection of massive particle. Throughout the paper, we mainly consider the case of $\lambda_2\geq0$, i.e., $\beta\leq0$ or $\beta\geq1/2$.

\subsection{Thermodynamic properties}
In this subsection, we briefly discuss the thermodynamic properties of BH in Eq.~\eqref{Kerrmetric}. First, the BH horizons are determined by
\begin{eqnarray}
\label{horizonsfunction}
&& \Delta(r)=r^2-2Mr+a^2+4a_0\lambda_1r^{2-\lambda_2}=0.
\end{eqnarray}
Obviously, when $\beta\rightarrow \pm\infty$, the horizons become
\begin{eqnarray}
&& r_{\pm}=M\pm\sqrt{M^2-a^2-\frac{a_0}{2}},
\end{eqnarray}
and other situations for $-\infty<\beta<+\infty$ are complicated. Fig.~\ref{fighorizon} shows the relationship between the BH horizon and parameters $a$, $a_0$ and $\beta$ in some special cases. Now, it is assumed that the BH has the outer horizon
and one can write the mass in terms of outer horizon $r_+$ from Eq.~\eqref{horizonsfunction}, as follows
\begin{eqnarray}
&& M=\frac{1}{2r_+}\left(r_+^2+a^2+4a_0\lambda_1r_+^{2-\lambda_2}\right).
\end{eqnarray}
Then, the tunneling method~\cite{Parikh2000,Banerjee2008,Sakti2019} can be used to calculate the Hawking temperature. In tunneling method, setting $d\theta=d\phi=0$ and $\theta=0$, the metric and Hawking temperature can be written as
\begin{eqnarray}
&&ds^2=-\hat f(r)dt^2+\frac{1}{\hat{f}(r)}dr^2,\\
&&T_\mathrm{H}=\frac{\partial_r \hat f(r)}{4\pi}\mid_{r=r_+}.
\end{eqnarray}
With this approach, our metric becomes
\begin{eqnarray}
&&ds^2=-\frac{\Delta}{r^2+a^2}dt^2+\frac{r^2+a^2}{\Delta}dr^2,
\end{eqnarray}
and the Hawking temperature reads
\begin{eqnarray}
\label{Hawking}
&&T_\mathrm{H}=\frac{\left(r_+-M\right)+2a_0r^{1-\lambda_2}\lambda_1\left(2-\lambda_2\right)}{2\pi\left(r_+^2+a^2\right)}~,
\end{eqnarray}
which can leads to the Hawking temperature of Kerr BH when $a_0=0$~\cite{Banerjee2008}. The angular velocity of metric in Eq.~\eqref{Kerrmetric} is
\begin{eqnarray}
&&\Omega=-\frac{g_{03}}{g_{33}}\mid_{r=r_+}=\frac{a}{r_+^2+a^2}.
\end{eqnarray}
The area of the BH is
\begin{eqnarray}
&&A_\mathrm{BH}=\int \sqrt{g_{22}~g_{33}}~d\theta d\phi=4 \pi \left(r_+^2+a^2\right),
\end{eqnarray}
and the Bekenstein-Hawking entropy of the BH can be calculated by area of the BH as follows
\begin{eqnarray}
&& S_\mathrm{BH}=\frac{A_\mathrm{BH}}{4}=\pi \left(r_+^2+a^2\right).
\end{eqnarray}
One can see that the string parameter $a_0$ has the effects to angular velocity and Bekenstein-Hawking entropy by affecting $r_+$. However, the expression of angular velocity and Bekenstein-Hawking entropy is the same with the Kerr BH, which is different from Kerr-Newman-NUT-Quintessence BH~\cite{Sakti2019}.

\begin{widetext}
\begin{center}
\begin{figure}[t]
\includegraphics[width=0.3\textwidth]{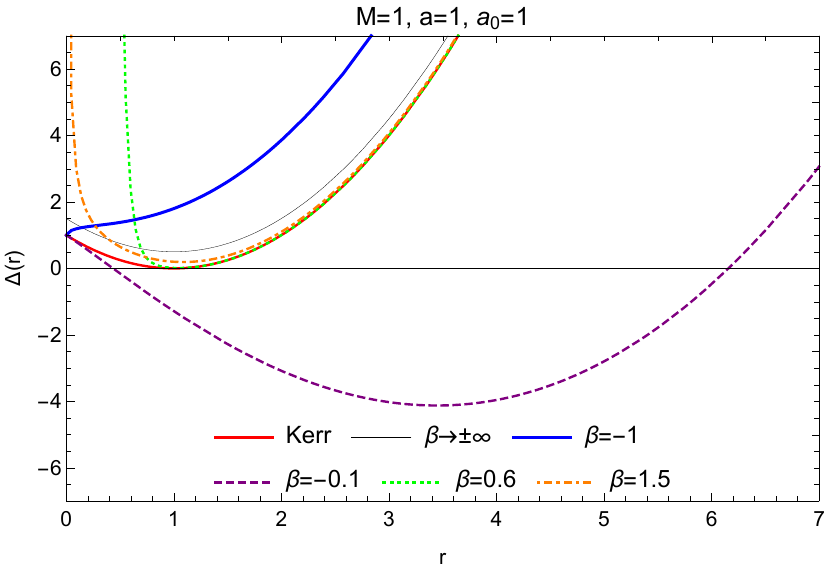}\hspace{0.04\textwidth}
\includegraphics[width=0.3\textwidth]{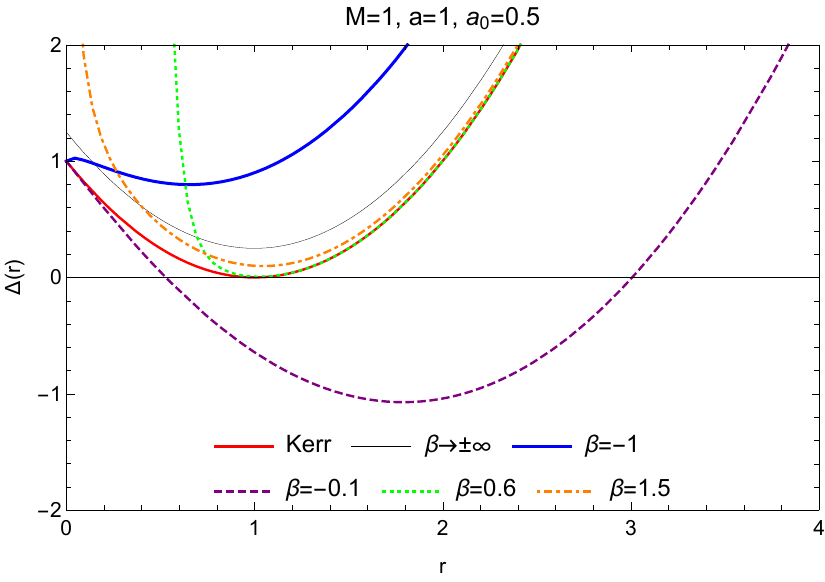}\hspace{0.04\textwidth}
\includegraphics[width=0.3\textwidth]{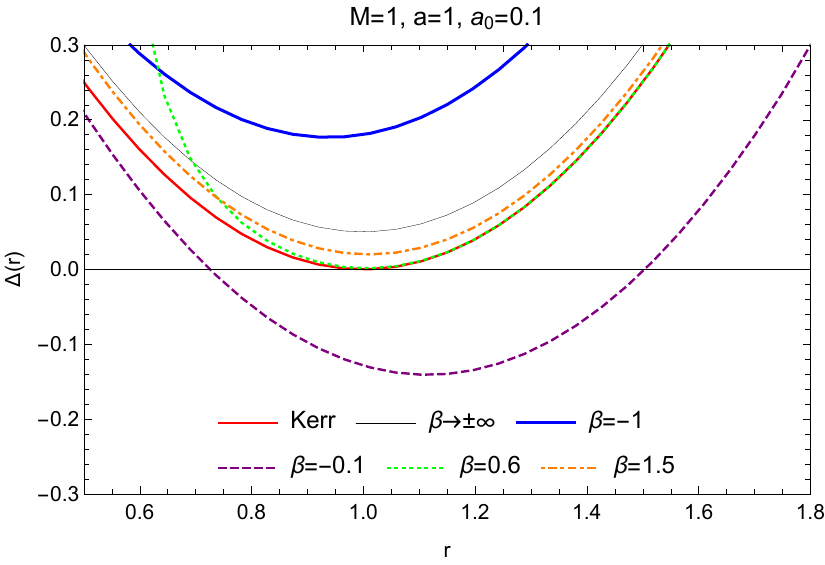}\\
(a)\hspace{0.3\textwidth}(b)\hspace{0.3\textwidth}(c)\\
\includegraphics[width=0.3\textwidth]{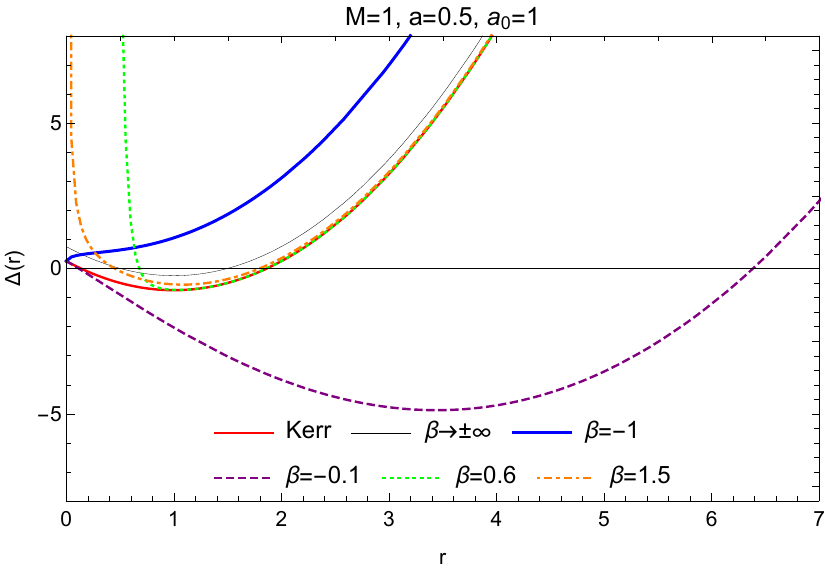}\hspace{0.04\textwidth}
\includegraphics[width=0.3\textwidth]{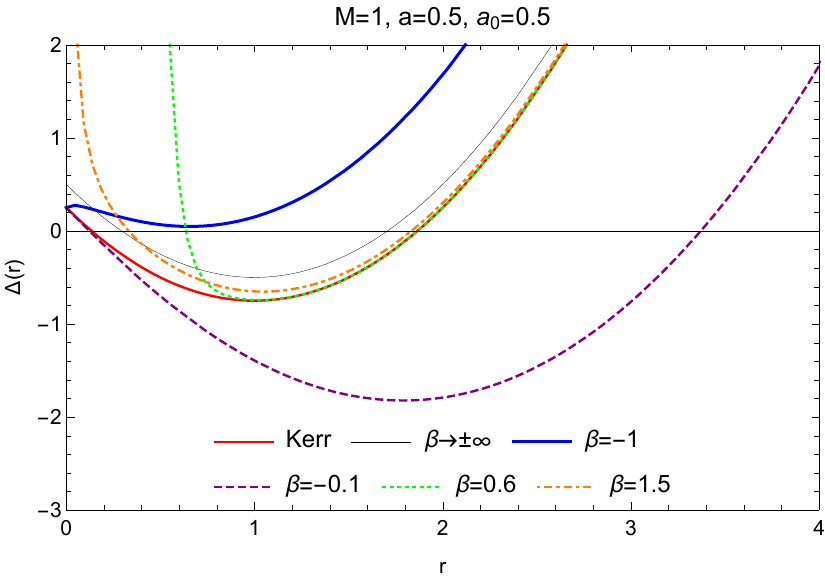}\hspace{0.04\textwidth}
\includegraphics[width=0.3\textwidth]{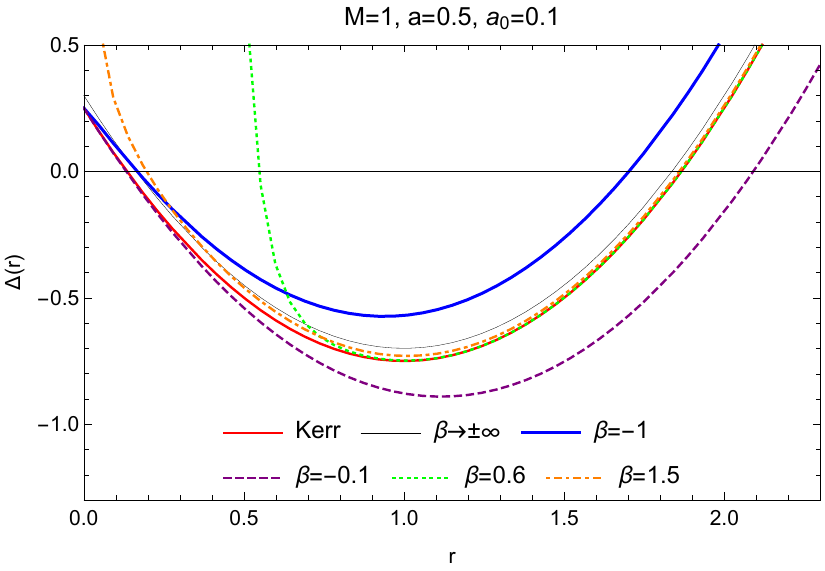}\\
(d)\hspace{0.3\textwidth}(e)\hspace{0.3\textwidth}(f)\\
\includegraphics[width=0.3\textwidth]{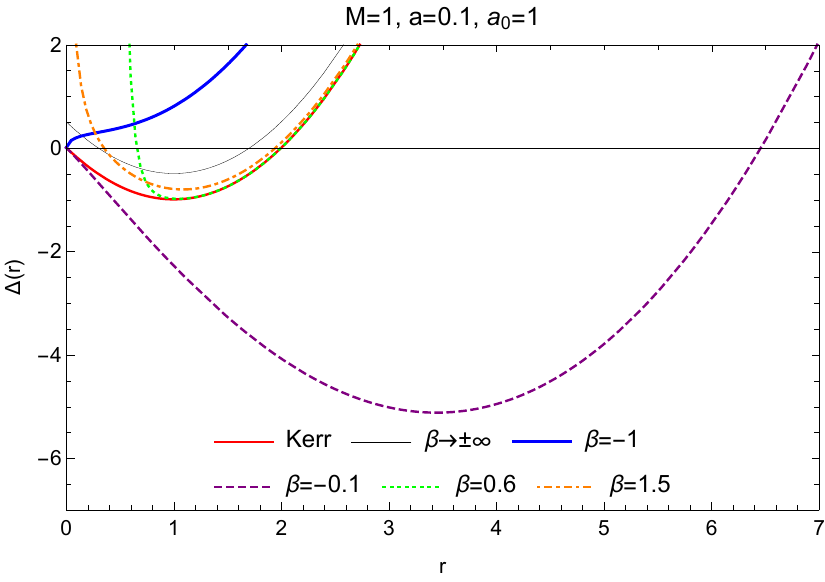}\hspace{0.04\textwidth}
\includegraphics[width=0.3\textwidth]{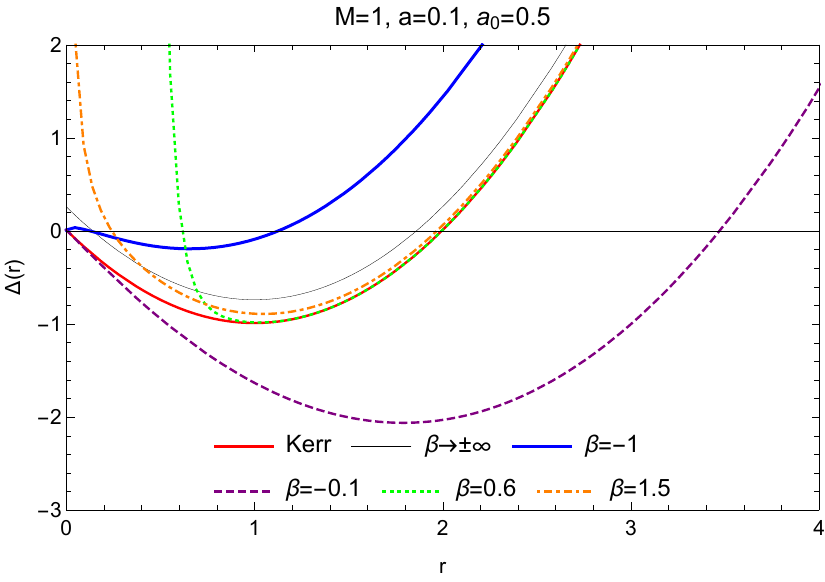}\hspace{0.04\textwidth}
\includegraphics[width=0.3\textwidth]{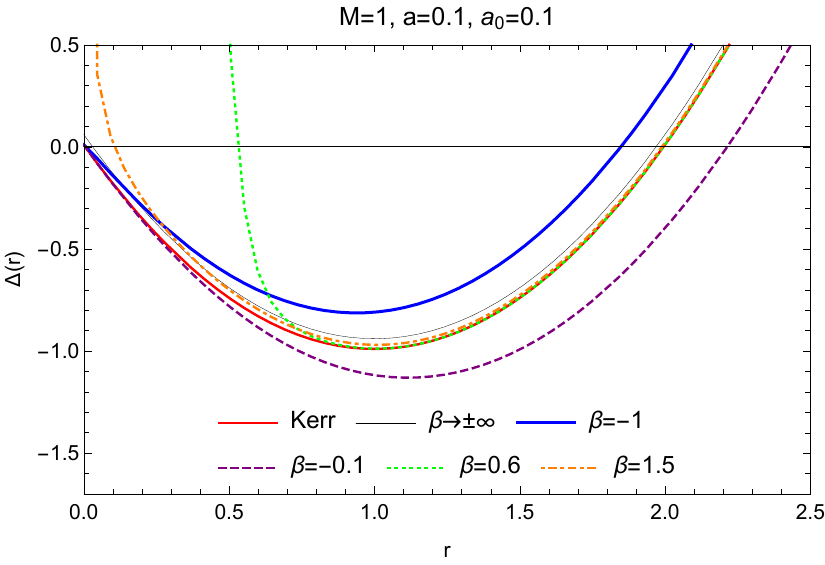}\\
(g)\hspace{0.3\textwidth}(h)\hspace{0.3\textwidth}(i)
\caption{The BH horizons, where vertical axis is denoted by $\Delta(r)$ and horizontal axis by $r$. We set $M=1$ and take $a$, $a_0$ and $\beta$ as variables.  For the pictures in first line, $a=1$ and $a_0$ takes values 1, 0.5 and 0.1 corresponding to (a), (b) and (c), respectively. For the pictures in the second line, $a=0.5$ and $a_0$ takes values 1, 0.5 and 0.1 corresponding to (d), (e) and (f), respectively. For the pictures in the last line, $a = 0.1$, and $a_0$ takes values 1, 0.5 and 0.1 corresponding to (g), (h) and (i), respectively. In all the nine figures, the thin black line, thick blue line, purple dashed line, green dotted line and orange dash-dotted line correspond to $\beta\rightarrow \pm\infty$, $\beta=-1$, $\beta=-0.1$, $\beta=0.6$ and $\beta=1$, respectively. Note that we have added Kerr BH ($a_0=0$) in each picture, plotted by red solid lines.} \label{fighorizon}
\end{figure}
\end{center}
\end{widetext}

\section{Infinite-distance gravitational deflection of massive particles}
\label{InGBDEF}

\subsection{JMRF metric}

The Jacobi (or Jacobi-Maupertuis) metric is a powerful tool in geometrodynamics. In 2016, Gibbons~\cite{Gibbons2016} derived the Jacobi metric of a static spacetime, which is an energy-dependent positive Riemannian metric. In 2019, Chanda \textit{et al.}~\cite{Chanda2019} discovered the Jacobi metric of stationary spacetime, which is an energy-dependent Randers-Finsler metric. A more general discussion about Jacobi metric can be found in Ref.~\cite{Chanda2019b}. The spatial part of time-like geodesic in spacetime is the spatial geodesic in the corresponding Jacobi metric space and thus the Jacobi metric method has been widely used in the study of particle motion in curved spacetime. For example, it was used to study Kepler orbit~\cite{Chanda2017a,Chanda2017b}, gravitational deflection of massive particles~\cite{LHZ2019,LJ2019,LA2019}, the motion of charged particles~\cite{Das2017}, and Hawking radiation~\cite{Bera2019}. As mentioned above, the Jacobi-Maupertuis metric of a stationary spacetime is a Randers-Finsler metric described by $(\bar{\alpha},\bar{\beta})$, say
\begin{eqnarray}
&&\mathcal{F}(x^i,dx^i)=ds_J=\sqrt{\bar{\alpha}_{ij}dx^idx^j}+\bar{\beta}_i dx^i,\nn
\end{eqnarray}
where $\bar{\alpha}_{ij}$ is a Riemannian metric and $\bar{\beta}_i $ is a one-form, satisfying $\bar\alpha^{ij}\bar\beta_i \bar\beta_j<1$. The Jacobi metric of a stationary spacetime, JMRF metric, reads~\cite{Chanda2019}
\begin{eqnarray}
&&\bar{\alpha}_{ij}=\frac{E^2+m^2g_{00}}{-g_{00}}\gamma_{ij}~,\\
&&\bar{\beta}_{i}=-E\frac{g_{0i}}{g_{00}}~,
\end{eqnarray}
where $m$ and $E$ is the mass and energy of the particle, respectively, and the spatial metric $\gamma_{ij}$ is defined by
\begin{eqnarray}
&&\gamma_{ij}\equiv g_{ij}-\frac{g_{0i}g_{0j}}{g_{00}}~.\nn
\end{eqnarray}
It should be noted that Crisnejo~\textit{et al.} recently derived JMRF metrics with optical media method~\cite{CGJ2019}. For metric in Eq.~\eqref{Kerrmetric}, one can obtain the corresponding JMRF metric as follows
\begin{eqnarray}
\label{KerrJMRE}
\bar\alpha_{ij}dx^i dx^j&=&\left(\frac{E^2}{1-\frac{A}{\Sigma}}-m^2\right)\bigg[\frac{\Sigma }{\Delta}dr^2+\Sigma d\theta^2+\sin^2\theta \nn\\
&&\times\bigg(a^2+r^2+\frac{a^2A\sin^2\theta\left(\Sigma-4A\right)}{\Sigma^2}\bigg) d\phi^2\bigg],\nn\\
\bar{\beta}_i dx^i &=& -\frac{aEA\sin^2\theta}{\Sigma-A}d\phi.
\end{eqnarray}
On the equatorial plane ($\theta=\pi/2$), the Randers-Finsler metric can be reduced to
\begin{eqnarray}
\label{EPKerrJMRE}
\bar\alpha_{ij}dx^i dx^j&=&\left(\frac{E^2}{1-\frac{A}{r^2}}-m^2\right)\bigg[\frac{r^2 }{\Delta}dr^2\nn\\
&&+\bigg(a^2+r^2+\frac{a^2A\left(r^2-4A\right)}{r^4}\bigg) d\phi^2\bigg],\nn\\
\bar{\beta}_i dx^i&=&-\frac{aEA}{r^2-A}d\phi.
\end{eqnarray}
The energy of the particle $E$ can be expressed by the velocity $v$ of the particle
\begin{eqnarray}
 E=\frac{m}{\sqrt{1-v^2}},
\end{eqnarray}
and this expression will be used in the following.

\subsection{Osculating Riemannian manifold and Gauss-Bonnet theorem}
Denoting a smooth manifold by $M$, the Finsler metric $\mathcal{F}(x,y)$ is a function on the tangent bundle $TM$, and its Hessian reads~\cite{Chern2002}
\begin{eqnarray}
\label{Hessian}
 g_{ij}(x,y)&=&\frac{1}{2}\frac{\partial^2\mathcal{F}^2(x,y)}{\partial y^i \partial y^j}.
\end{eqnarray}
To introduce the GB theorem to study the light deflection, Werner~\cite{Werner2012} applied Naz{\i}m's method to construct an osculating Riemannian manifold $(M, \tilde{g})$ of Finsler manifold $(M, \mathcal{F})$. Following this scheme, a smooth nonzero vector field $Y$ which is tangent to the geodesic $\eta_{\mathcal{F}}$, say $Y(\eta_{\mathcal{F}})=y$, can be chosen, and a Riemannian metric can be obtained by Hessian
\begin{eqnarray}
\label{Hessian1}
 \tilde{g}_{ij}(x)&=&g_{ij}\left(x,Y(x)\right).
\end{eqnarray}
By this construction, the geodesic in $(M, \mathcal{F})$ is still a geodesic in $(M, \tilde{g})$, i.e. $\eta_{\mathcal{F}}=\eta_{\tilde{g}}$, for which the details are shown in Ref.~\cite{Werner2012}. On the equatorial plane, our Randers-Finsler metric in Eq.~\eqref{EPKerrJMRE} leads to
\begin{widetext}
\begin{eqnarray}
\label{Randers-FinslerY}
\mathcal{F}\left(r,\phi,Y^r,Y^\phi\right)&=&\sqrt{m^2\left(\frac{1}{\left(1-\frac{A}{r^2}\right)\left(1-v^2\right)}-1\right)\bigg[\frac{r^2 }{\Delta}(Y^r)^2+\bigg(a^2+r^2+\frac{a^2A\left(r^2-4A\right)}{r^4}\bigg) (Y^\phi)^2\bigg]}-\frac{amA}{\left(r^2-A\right)\sqrt{1-v^2}}Y^{\phi}~.~~~~~~~~
\end{eqnarray}

In the present work, we mainly consider the terms including $M, M^2, M a, a^2, a_0$ in deflection angle and the terms containing $M a_0, a a_0$ are ignored. To simplify writing, the approximation is denoted as $\mathcal O(\varepsilon^3) \equiv\mathcal O(M^3,M^2a,Ma^2,a^3,a_0^2,Ma_0,aa_0)$. To this end, the first-order particle ray will be considered, which is consistent with Kerr spacetime~\cite{LJ2019}
\begin{eqnarray}
\label{particleray}
r(\phi)&=&\frac{b}{\sin \phi}-\left(\cot^2 \phi+\frac{\csc^2 \phi}{v^{2}}\right)M+\mathcal O(M^2,a^2,Ma,a_0).
\end{eqnarray}
But now, we only need the zero-order particle ray $r=b/\sin \phi$ to construct the following vector fields
\begin{eqnarray}
\label{vector field}
&&Y^r=\frac{dr}{dl}=-\frac{\cos\phi}{m v}\sqrt{1-v^2}~,\nn\\
&&Y^\phi=\frac{d\phi}{dl}=\frac{\sin^2\phi}{m b v}\sqrt{1-v^2}.
\end{eqnarray}
Then, substituting Eq.~\eqref{Randers-FinslerY} into Eq.~\eqref{Hessian}, and using Eq.~\eqref{vector field}, the osculating Riemannian metric can be obtained as follows

\begin{eqnarray}
\tilde{g}_{rr}&=&\frac{m^2}{1-v^2}\bigg[v^2+\frac{2m\left(1+v^2\right)}{r}-\frac{2a M r v\sin^6\phi}{b^3\left(\cos^2\phi+\frac{r^2}{b^2}\sin^4\phi\right)^{3/2}}\nn\\
&&-4a_0r^{-\lambda_2}(1+v^2)\lambda_1+\frac{4\left(2+v^2\right)M^2}{r^2}-\frac{a^2v^2}{r^2}\bigg]+\mathcal O(\varepsilon^3),\\
\tilde{g}_{r\phi}&=&\frac{m^2}{1-v^2}\frac{2M av\cos^3\phi}{r\left(\cos^2\phi+\frac{r^2}{b^2}\sin^4\phi\right)^{3/2}}+\mathcal O(\varepsilon^3),\\
\tilde{g}_{\phi\phi}&=&\frac{m^2}{1-v^2}\bigg[v^2 r^2+2 M r-\frac{2M a r v  \sin^2\phi\left(3\cos^2\phi+2\frac{r^2}{b^2}\sin^4\phi\right)}{b\left(\cos^2\phi+\frac{r^2}{b^2}\sin^4\phi\right)^{3/2}}\nn\\
&&-4a_0r^{2-\lambda_2}\lambda_1+4M^2+a^2v^2\bigg]+\mathcal O(\varepsilon^3).
\end{eqnarray}
\end{widetext}

Now, the particle ray is a geodesic in the osculating-Riemannian-metric spacetime. In the next, we will use the GB theorem to study the gravitational deflection, and let us briefly introduce the GB theorem first. The GB theorem deeply connects the geometry and topology of a surface. Let $D$ be a compact, oriented surface with Gaussian curvature $\mathcal{K}$ and Euler characteristic $\chi(D)$, and its boundary $\partial{D}$ is a piecewise smooth curve with geodesic curvature $k$. The GB theorem states that~\cite{GW2008,Carmo1976}
\begin{equation}
\iint_D{\mathcal{K}}dS+\oint_{\partial{D}}k~dl+\sum_{i=1}{\varphi_i}=2\pi\chi(D),\\
\end{equation}
where $dS$ is the area element, $dl$ is the line element of the boundary, and $\varphi_i$ is the jump angle in the $i$-th vertex of $\partial{D}$ in the positive sense, respectively.

\begin{figure}[t]
\label{Figure}
\centering
\includegraphics[width=8.0cm]{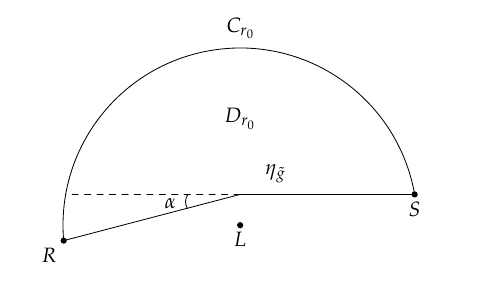}
\caption{A region $D_{r_0}\subset(M,\tilde{g}_{ij})$, with $\partial D_{r_0}=\eta_{\tilde{g}} \cup C_{r_0}$. The particle ray $\eta_{\tilde{g}}$ is a geodesic in $D_{r_0}$ and the curve $C_{r_0}$ is defined by $r(\phi)=r_0=\mathrm{constant}$. $S$, $R$ and $L$ denote the source, the receiver, and the lens, respectively. $\alpha$ is the deflection angle. Note that $\varphi_S+\varphi_R\rightarrow\pi$ as $r_0\rightarrow \infty$.}
\end{figure}

Then, one can consider a region $D_{r_0}\subset(M,\tilde{g}_{ij})$ bounded by a particle ray $\eta_{\tilde{g}}$ from the source $S$ to the receiver $R$ and a curve $C_{r_0}$ defined by $r=r_0=\mathrm{constant}$, i.e., $\partial D_{r_0}=\eta_{\tilde{g}} \cup C_{r_0}$. Due to the fact that $\eta_{\tilde{g}}$ is a geodesic in $D_{r_0}$, one can have $k(\eta_{\tilde{g}})=0$. For the curve $C_{r_0}$, when $r_0\rightarrow \infty$, we have $k(C_{r_0})dl\rightarrow d\phi$, and therefore $\int_{C_{\infty}}k(C_{\infty}) dl=\int_0^{\pi+\alpha|_{\infty}}d\phi$, with $\alpha|_{\infty}$ the infinite-distance deflection angle. In addition, one can see $\chi(D_{r_0})=1$ in the region outside of the lens object $L$, and noticing that $\varphi_R+\varphi_S\rightarrow\pi$ as $r_0\rightarrow\infty$, finally one can have
\begin{eqnarray}
&&\iint_{D_{r_0}}\tilde{\mathcal{K} }dS+\int_S^Rd\phi+\varphi_R+\varphi_S\nn\\
&&\stackrel{r_0\rightarrow\infty}{=}\iint_{D_{\infty}}\tilde{\mathcal{K} } dS+\int_0^{\pi+\alpha|_{\infty}}d\phi+\pi=2\pi,
\end{eqnarray}
where GB theorem has been applied to the region $D_{r_0}$. From this, the infinite-distance deflection of massive particles can be calculated by
\begin{eqnarray}
\label{INGBangle}
\alpha|_{\infty}&=&-\iint_{D_\infty}\tilde{\mathcal{K} } dS.
\end{eqnarray}
In the expression above, one can see that the deflection angle is coordinate-invariant, and additionally, the Gaussian curvature of a surface defined by Remiannian metric $g_{ij}$ in equatorial plane $(r,\phi)$ can be calculated as in Ref.~\cite{Werner2012}
\begin{eqnarray}
\label{Gauss-K}
\mathcal{K}&=&\frac{1}{\sqrt{\det g}}\left[\frac{\partial \left(\frac{\sqrt{\det g}}{g_{rr}}{{\Gamma}^\phi_{rr}}\right)}{\partial{\phi}}-\frac{\partial\left(\frac{\sqrt{\det g}}{g_{rr}}{{\Gamma}^\phi_{r\phi}}\right)}{\partial{r}}\right].
\end{eqnarray}

\subsection{Infinite-distance deflection angle of massive particles}
Now, according to Eq.~\eqref{INGBangle}, we can investigate the effects of a cloud of strings on gravitational deflection angle of massive particles for a receiver and source at infinite distance from the rotating BH in Rastall gravity. First, the Gaussian curvature of osculating Riemannian metric $\tilde g_{ij}$ can be calculated by Eq.~\eqref{Gauss-K} and the result is
\begin{eqnarray}
\label{KKerrGausss}
\tilde{\mathcal{K} }&=&-\frac{\left(1-v^4\right)M}{m^2 r^3v^4}+\frac{2a_0\left(1-v^2\right)\lambda_1 \lambda_2\left(v^2+\lambda_2\right)}{m^2v^4r^{2+\lambda_2}}\nn\\
&&+\frac{3\left(2-3v^2+v^4\right)M^2}{m^2r^4v^6}+\frac{3M a\left(1-v^2\right)} {4m^2 v^3 b^2 r^2 } \Xi\left(r,\phi\right)+\mathcal O(\varepsilon^3),\nn\\
\end{eqnarray}
where
\begin{eqnarray}
\Xi\left(r,\phi\right)&=&\frac{\sin^3\phi}{\left(\cos^2\varphi+\frac{r^2}{b^2}\sin^4\phi\right)^{7/2}}\Bigg[2\cos^6\phi\left(-2+\frac{5r}{b}\sin\phi\right)\nn\\
&&+\frac{2r}{b}\cos^2\phi\sin^5\phi\left(2-\frac{r^2}{b^2}+\frac{r^2}{b^2}\cos2\phi+\frac{4r}{b}\sin\phi\right)\nn\\
&&+\cos^4\phi\sin^2\phi\left(-2+\frac{9r}{b}\sin\phi-\frac{10r^3}{b^3}\sin^3\phi\right)\nn\\
&&+\frac{r^2}{b^2}\left(-\frac{r}{b}\sin^9\phi+\frac{2r^3}{b^3}\sin^{11}\phi+\sin^42\phi\right)\Bigg].
\end{eqnarray}
It can be found that there is no $a^2$ term in Gaussian curvature. Then, by Eq.~\eqref{INGBangle}, the infinite-distance deflection angle of massive particles can be obtained in the following
\begin{eqnarray}
\label{Kerrangle1}
\alpha|_{\infty}&=&-\int_0^{\pi}\int_{r(\phi)}^\infty \tilde{\mathcal{K} }\sqrt{\det\tilde{g}}~dr d\phi\nn\\
&=&\int_0^{\pi}\int_{r(\phi)}^\infty \bigg[\frac{\left(1+v^2\right)M}{r^2v^2}-\frac{2a_0\lambda_1 \lambda_2\left(v^2+\lambda_2\right)}{v^2r^{1+\lambda_2}}\nn\\
&&+\frac{(6v^2+v^4-4)M^2}{r^3v^4}-\frac{3M a} {r v b^2} \Xi\left(r,\phi\right)\bigg]dr d\varphi\nn\\
&&+\mathcal O(\varepsilon^3)\nn\\
&=&\alpha_{Kerr}|_{\infty}+\alpha_{a_0}|_{\infty}+\mathcal O(\varepsilon^3),
\end{eqnarray}
where $r(\phi)$ is the first-order particle ray in Eq.~\eqref{particleray}. In the above, $\alpha_{Kerr}|_{\infty}$ denotes the infinite-distance angle of massive particles in Kerr spacetime, and $\alpha_{a_0}|_{\infty}$ is the part of deflection angle related to a cloud of strings,
\begin{eqnarray}
\label{Kerrangle2}
\alpha_{Kerr}|_{\infty}&\equiv&\int_0^{\pi}\int_{b/{\sin\phi}}^\infty \bigg[\frac{\left(1+v^2\right)M}{r^2v^2}+\frac{(6v^2+v^4-4)M^2}{r^3v^4}\nn\\
&&-\frac{3M a} {r v b^2} \Xi\left(r,\phi\right)\bigg]dr d\varphi\nn\\
&=&\frac{2M\left(1+v^2\right)}{bv^2}-\frac{4M a}{b^2v}\nn,\\
\alpha_{a_0}|_{\infty}&\equiv&-\int_0^{\pi}\int_{b/{\sin\phi}}^\infty \frac{2a_0\lambda_1 \lambda_2\left(v^2+\lambda_2\right)}{v^2r^{1+\lambda_2}}drd\phi.
\end{eqnarray}
When $a_0=0$, we have $\alpha_{a_0}|_{\infty}=0$, and the result agrees with the deflection angle of massive particles in Kerr spacetime~\cite{He2017b}. Notice that we have assumed that $\lambda_2\geq0$. Therefore, another expression of $\alpha_{a_0}|_{\infty}$ in Eq.~\eqref{Kerrangle2} can be derived, and the result is expressed by gamma function
\begin{eqnarray}
\alpha_{a_0}|_{\infty}&=&-\frac{2a_0\sqrt{\pi}\lambda_1\left(\lambda_2+v^2\right)}{b^{\lambda_2}v^2}\frac{\Gamma[\frac{1}{2}+\frac{\lambda_2}{2}]}{\Gamma[1+\frac{\lambda_2}{2}]}.
\end{eqnarray}
Finally, Eq.~\eqref{Kerrangle1} can be rewritten as
\begin{eqnarray}
\label{infangle}
\alpha|_{\infty}&=&
\frac{2M\left(1+v^2\right)}{bv^2}-\frac{2a_0\sqrt{\pi}\lambda_1\left(\lambda_2+v^2\right)}{b^{\lambda_2}v^2}\frac{\Gamma[\frac{1}{2}+\frac{\lambda_2}{2}]}{\Gamma[1+\frac{\lambda_2}{2}]}\nn\\
&&+\frac{3\pi \left(4+v^2\right)M^2}{4b^2 v^2}\pm\frac{4M a}{b^2v}+\mathcal O(\varepsilon^3),
\end{eqnarray}
where positive and negative signs correspond to retrograde and prograde particle orbits, respectively. For light case ($v=1$), the deflection angle reads
\begin{eqnarray}
\alpha|_{\infty}(v=1)&=&
\frac{4M}{b}-\frac{2a_0\sqrt{\pi}\lambda_1\left(\lambda_2+1\right)}{b^{\lambda_2}}\frac{\Gamma[\frac{1}{2}+\frac{\lambda_2}{2}]}{\Gamma[1+\frac{\lambda_2}{2}]}\nn\\
&&+\frac{15\pi M^2}{4b^2}\pm\frac{4M a}{b^2}+\mathcal O(\varepsilon^3).
\end{eqnarray}
one can see that the effect of a cloud of strings on infinite-distance deflection angle of light has not been eliminated.

\section{Finite-distance gravitational deflection of massive particles}\label{GBDEF}
In the previous section, with Werner's method, we studied the infinite-distance gravitational deflection of massive particles in rotating BH surrounded by a cloud of strings. In this section, we will consider the finite-distance deflection, a more general situation where the receiver and the source at finite distance from the lens object. In this case, Werner's method seems not feasible, and Ono \textit{et al} proposed a generalized optical method to study the finite-distance deflection of light~\cite{OIA2017}. Recently, Li~\textit{et al.}~\cite{LJ2019,LA2019} extended the study to massive particles case based on JMRF metric.

\subsection{The generalized Jacobi metric method}
Here, it is assumed that the particles move in a 3-dimensional Riemannian space $\bar{M}$ defined by generalized Jacobi metric $\bar{\alpha}_{ij}$ in Eq.~\eqref{KerrJMRE},
\begin{eqnarray}
\label{GJM}
&& dl^2=\bar{\alpha}_{ij}dx^idx^j.
\end{eqnarray}
Now, the motion of particles does not follow the geodesic in $\bar{M}$, and then Eq.~\eqref{INGBangle} is not available. However, by the GB theorem one can add a term related to geodesic along particle ray to calculate the gravitational deflection angle. For metric in Eq.~\eqref{KerrJMRE}, the geodesic curvature along particle ray $\eta_{\bar{\alpha}}$ can be calculated~\cite{OIA2017}
\begin{eqnarray}
\label{geodesic}
&& k (\eta_{\bar{\alpha}})=-\frac{\frac{\partial \bar{\beta}_{\varphi}}{\partial r}}{\sqrt{{\det{\bar{\alpha}}~\bar{\alpha}^{\theta\theta}}}}.
\end{eqnarray}

To study the finite-distance deflection, we apply the definition of deflection angle proposed in Ref.~\cite{OIA2017}
\begin{eqnarray}
\label{angle}
&& \alpha\equiv \Psi_R-\Psi_S+\phi_{RS},
\end{eqnarray}
where $\Psi_R$ and $\Psi_S$ are angles between the tangent of the particle ray and the radial direction from the lens to receiver and source, respectively, and the coordinate angle $\phi_{RS}\equiv\phi_R-\phi_S$.

\begin{figure}[t]
\centering
\includegraphics[width=8.5cm]{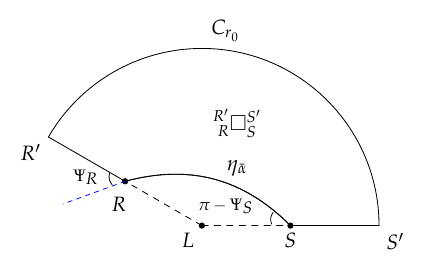}
\caption{The quadrilateral $\prescript{R'}{R}\Box_{S}^{S'}\subset(\bar{M},\alpha_{ij})$. $\eta_{\bar{\alpha}}$ is the particle ray from source $S$ to receiver $R$, deflected by lens object $L$. $R'$ and $S'$ are the intersection points of curve $C_{r_0}$ and radial directions, from the $L$ to $R$ and $S$, respectively. $\Psi_R$ and $\Psi_S$ are angles between the tangent of the particle ray and the radial direction from the lens to $R$ and $S$, respectively. $\varphi_R=\Psi_R$ and $\varphi_S=\pi-\Psi_S$, and in addition, $\varphi_{R'}+\varphi_{S'}\rightarrow \pi$ as $r_0\rightarrow\infty$.}\label{Figure2}
\end{figure}

Let us now consider the quadrilateral $\prescript{R'}{R}\Box_{S}^{S'}\subset(\bar{M},\bar{\alpha}_{ij})$ with Gaussian curvature $\bar{\mathcal{K}}$, as shown in Fig.~\ref{Figure2}. It is bounded by four curves: the particle ray $\eta_{\bar{\alpha}}$, a curve $C_{r_0}$ as defined in the previous section, two spatial geodesics of outgoing radial lines from $R$ to $R'$, and $S$ to $S'$, respectively. For this region, one can see $\chi(\prescript{R'}{R}\Box_{S}^{S'})=1$, and additionally, we have $\varphi_S=\pi-\Psi_S$ and $\varphi_R=\Psi_R$. Thus, using GB theorem to the quadrilateral, one can obtain

\begin{eqnarray}
\label{gbangle}
&&\iint_{_{R}^{R'}\Box_{S}^{S'}}\bar{\mathcal{K}} dS-\int_{S}^{R}kdl+\int_{C_{r_0}}kdl\nn\\
&&+\Psi_R+\left(\pi-\Psi_S\right)+\varphi_{S'}+\varphi_{R'}=2\pi.
\end{eqnarray}
For curve $C_{r_0}$, $\int_{C_{\infty}}k(C_\infty) dl=\phi_{RS}$, and furthermore, the sum of two jump angles $\varphi_{R'}+\varphi_{S'}=\pi$ as $r_0\rightarrow\infty$. Therefore, when $r_0\rightarrow \infty$, Eq.~\eqref{gbangle} leads to
\begin{eqnarray}
\label{angle1}
\iint_{_{R}^{\infty}\Box_{S}^{\infty}}\bar{\mathcal{K}} dS-\int_{S}^{R}k(\eta_{\bar{\alpha}})dl+\phi_{RS}+\Psi_R-\Psi_S=0.
\end{eqnarray}
Then, according to the definition of deflection angle in Eq.~\eqref{angle}, the above equation can be rewritten as
\begin{eqnarray}
\label{GBangle}
\alpha&=&-\iint_{_{R}^{\infty}\Box_{S}^{\infty}}\bar{\mathcal{K}} dS+\int_{S}^{R}k(\eta_{\bar{\alpha}}) dl.
\end{eqnarray}
This expression shows that the finite-distance deflection angle is also coordinate-invariant. In addition, this expression can be used to calculate the infinite-distance deflection as long as we let $r_R\rightarrow \infty$ and $r_S\rightarrow \infty$, and this is the reason we say that the finite-distance deflection is more common than infinite-distance deflection.

\subsection{Finite-distance deflection angle of massive particles}
Substituting the corresponding metric components in Eq.~\eqref{EPKerrJMRE} into Eq.~\eqref{Gauss-K}, one can obtain the Gaussian curvature of the generalized Jacobi metric as follows
\begin{eqnarray}
\label{KerrGauss}
\bar{\mathcal{K}}&=&\frac{2a_0\lambda_1\lambda_2\left(1-v^2\right)\left(\lambda_2+v^2\right)}{m^2v^4r^{2+\lambda_2}}-\frac{\left(1-v^4\right)M}{m^2v^4r^3}\nn\\
&&+\frac{3\left(2-3v^2+v^4\right)M^2}{m^2r^4v^6}+\mathcal O(\varepsilon^3)
\end{eqnarray}
On the other hand, the geodesic curvature of a particle ray in generalized Jacobi space $(\bar{M},\bar{\alpha}_{ij})$ can be calculated by Eq.~\eqref{geodesic}, and the result is
\begin{eqnarray}
\label{Kerrgeodesic}
k(\eta_{\bar{\alpha}})&=&-\frac{2\sqrt{1-v^2}M a}{mv^2r^3}+\mathcal O(\varepsilon^3).
\end{eqnarray}
From Eq.~\eqref{GJM}, one can have
\begin{eqnarray}
k(\eta_{\bar{\alpha}})dl&=&-\frac{2M a\sin\phi}{b^2v}d\phi+\mathcal O(\varepsilon^3)
\end{eqnarray}
where we have used the first-order particle ray in Eq.~\eqref{particleray}. Finally, substituting the quantities mentioned above into Eq.~\eqref{GBangle}, the deflection angle is
\begin{eqnarray}
\label{Kerrangle}
\alpha&=&\int_{\phi_S}^{\phi_R}\int_{r(\phi)}^{\infty}\bigg[-\frac{2a_0\lambda_1\lambda_2\left(\lambda_2+v^2\right)}{v^2r^{1+\lambda_2}}+\frac{\left(1+v^2\right)M}{v^2r^2}\nn\\
&&+\frac{(6v^2+v^4-4)M^2}{r^3v^4}\bigg] drd\phi\nn\\
&&-\int_{\phi_S}^{\phi_R}\frac{2M a\sin\phi}{b^2v}d\phi+\mathcal O(\varepsilon^3).
\end{eqnarray}
According to the first-order particle ray in Eq.~\eqref{particleray}, one can obtain the first-order coordinate angle at receiver and source respectively,
\begin{eqnarray}
 \phi_S&=&\arcsin\left(\frac{b}{r_S}\right)+\frac{b^{2} v^{2}-r_S^{2}\left(1+v^{2}\right)}{r_S \sqrt{r_S^{2}-b^{2}} v^{2}}\frac{M}{b},\nn\\
 \phi_R&=&\pi-\arcsin\left(\frac{b}{r_R}\right)-\frac{b^{2} v^{2}-r_R^{2}\left(1+v^{2}\right)}{r_R \sqrt{r_R^{2}-b^{2}} v^{2}}\frac{M}{b}.
 \end{eqnarray}

When $\alpha_0=0$, Eq.~\eqref{Kerrangle} reduces to Kerr BH case, and the result becomes~\cite{LJ2019}
\begin{eqnarray}
\alpha_{Kerr}&=&\frac{\left(1+v^2\right)M\left(\sqrt{1-\frac{b^2}{r_R^2}}+\sqrt{1-\frac{b^2}{r_S^2}}\right)}{b v^2}\nn\\
&&\frac{3\left(4+v^2\right)\left[\pi-\arcsin(\frac{b}{r_R})-\arcsin(\frac{b}{r_S})\right]}{4v^2}\frac{M^2}{b^2}\nn\\
&&+\frac{\frac{b}{r_S}\left[3v^2\left(4+v^2\right)+\left(4-8v^2-3v^4\right)\frac{b^2}{r_S^2}\right]}{4v^4\sqrt{1-\frac{b^2}{r_S^2}}}\frac{M^2}{b^2}\nn\\
&&+\frac{\frac{b}{r_R}\left[3v^2\left(4+v^2\right)+\left(4-8v^2-3v^4\right)\frac{b^2}{r_R^2}\right]}{4v^4\sqrt{1-\frac{b^2}{r_R^2}}}\frac{M^2}{b^2}\nn\\
&&\pm\frac{2Ma\left(\sqrt{1-\frac{b^2}{r_R^2}}+\sqrt{1-\frac{b^2}{r_S^2}}\right)}{b^2v},
\end{eqnarray}
where the positive and negative signs correspond to retrograde and prograde particle orbits, respectively. Accordingly, the finite-distance deflection angle of massive particles in Eq.~\eqref{Kerrangle} can be rewritten as
\begin{eqnarray}
\label{KKerrangle}
\alpha&=&\alpha_{Kerr}+\alpha_{a_0}+\mathcal O(\varepsilon^3),
\end{eqnarray}
where
\begin{eqnarray}
\label{KKKerrangle}
\alpha_{a_0}&=&-2a_0\lambda_1\lambda_2\left(1+\frac{\lambda_2}{v^2}\right)\int_{\phi_S}^{\phi_R}\int_{\frac{b}{\sin\phi}}^{\infty}r^{-1-\lambda_2}drd\phi~,\nn\\
&=&-\frac{2a_0\lambda_1\left(\lambda_2+v^2\right)}{b^{\lambda_2}v^2}\int_{\phi_S}^{\phi_R}\sin^{\lambda_2}\phi d\phi\nn\\
&=&-\frac{2a_0\lambda_1\left(\lambda_2+v^2\right)}{b^{\lambda_2}v^2}\nn\\
&&\times\Bigg[\sqrt{1-\frac{b^2}{r_R^2}}{}_2F_1\bigg(\frac{1}{2},\frac{1-\lambda_2}{2};\frac{3}{2};1-\frac{b^2}{r_R^2}\bigg)\nn\\
&&+\sqrt{1-\frac{b^2}{r_S^2}}{}_2F_1\bigg(\frac{1}{2},\frac{1-\lambda_2}{2};\frac{3}{2};1-\frac{b^2}{r_S^2}\bigg)\Bigg].
\end{eqnarray}
Here, ${}_2F_1(a,b;c;z)$ is the Hypergeometric function~\cite{Bealsbook}. Let $v=1$, it is shown that the finite-distance deflection angle of light is also effected by the cloud of strings. In addition, for $r_R\rightarrow\infty$ and $r_S\rightarrow\infty$, the infinite-distance deflection angle of massive particles in Eq.~\eqref{infangle} can be recovered. Furthermore, the difference between the finite- and infinite-distance deflection angles can be described by the finite-distance correlation
\begin{eqnarray}
\delta \alpha=\alpha|_{\infty}-\alpha=\delta \alpha_{Kerr}+\delta \alpha_{a_0},
\end{eqnarray}
where $\delta \alpha_{Kerr}=\alpha_{Kerr}|_{\infty}-\alpha_{Kerr}$ and $\delta \alpha_{a_0}=\alpha_{a_0}|_{\infty}-\alpha_{a_0}$.

In the following, several specific values of $\lambda_2$ will be considered to show the effects of a cloud of strings on finite-distance gravitational deflection angle of massive particles.

\subsubsection{$\beta\rightarrow \pm\infty,\lambda_1=1/8,\lambda_2=2$}
When $\beta\rightarrow \pm\infty$, we have $\lambda_1=1/8,\lambda_2=2$, and Eq.~\eqref{KKKerrangle} becomes
\begin{eqnarray}
\alpha_{a_0}&=&-\frac{a_0\left(2+v^2\right)}{8b^2v^2}\Bigg(\frac{b}{r_R}\sqrt{1-\frac{b^2}{r_R^2}}+\frac{b}{r_S}\sqrt{1-\frac{b^2}{r_S^2}}\nn\\
&&+\pi-\arcsin\frac{b}{r_R}-\arcsin \frac{b}{r_S}\Bigg).
\end{eqnarray}
Meanwhile, the infinite-distance deflection angle can be recovered
\begin{eqnarray}
\alpha_{a_0}|_{\infty}&=&-\frac{a_0\pi\left(2+v^2\right)}{8b^2v^2}.
\end{eqnarray}

\subsubsection{$\beta=3/2,\lambda_1=1/20,\lambda_2=3$}
When $\beta=3/2$, we have $\lambda_1=1/20,\lambda_2=3$, and Eq.~\eqref{KKKerrangle} leads to
\begin{eqnarray}
\label{string1}
\alpha_{a_0}&=&-\frac{a_0\left(3+v^2\right)}{30b^3v^2}\Bigg[2\sqrt{1-\frac{b^2}{r_S^2}}+2\sqrt{1-\frac{b^2}{r_R^2}}\nn\\
&&+\frac{b^2\sqrt{1-\frac{b^2}{r_R^2}}}{r_R^2}+\frac{b^2\sqrt{1-\frac{b^2}{r_S^2}}}{r_S^2}\Bigg].
\end{eqnarray}
When $r_R\rightarrow\infty$ and $r_S\rightarrow\infty$, the infinite-distance deflection angle becomes
\begin{eqnarray}
\label{Istring1}
\alpha_{a_0}|_{\infty}&=&-\frac{2a_0\left(3+v^2\right)}{15b^3v^2}.
\end{eqnarray}

\subsubsection{$\beta=1,\lambda_1=1/36,\lambda_2=4$}
Let $\beta=1$, and we have $\lambda_1=1/36,\lambda_2=4$. In this case, Eq.~\eqref{KKKerrangle} comes to
\begin{eqnarray}
\label{string2}
\hat{\alpha}_{a_0}&=&-\frac{a_0\left(4+v^2\right)}{144b^4v^2}\Bigg[\frac{3b\sqrt{1-\frac{b^2}{r_R^2}}}{r_R}+\frac{3b\sqrt{1-\frac{b^2}{r_S^2}}}{r_S}\nn\\
&&+\frac{2b^3\sqrt{1-\frac{b^2}{r_R^2}}}{r_R^3}+\frac{2b^3\sqrt{1-\frac{b^2}{r_S^3}}}{r_S^3}\nn\\
&&+3\left(\pi-\arcsin\frac{b}{r_R}-\arcsin\frac{b}{r_S}\right)\Bigg].
\end{eqnarray}
For infinite-distance deflection
\begin{eqnarray}
\label{Istring2}
\alpha_{a_0}|_{\infty}=-\frac{a_0\pi\left(4+v^2\right)}{48b^4v^2}.
\end{eqnarray}

\subsubsection{$\beta=5/6,\lambda_1=1/56,\lambda_2=5$}
We consider $\beta=5/6$, which leads to $\lambda_1=1/56$ and $\lambda_2=5$, and in this case, Eq.~\eqref{KKKerrangle} becomes
\begin{eqnarray}
\label{string3}
\hat{\alpha}_{a_0}&=&-\frac{a_0\left(5+v^2\right)}{420b^5v^2}\Bigg[8\sqrt{1-\frac{b^2}{r_R^2}}+8\sqrt{1-\frac{b^2}{r_S^2}}\nn\\
&&+\frac{4b^2\sqrt{1-\frac{b^2}{r_R^2}}}{r_R^2}+\frac{4b^2\sqrt{1-\frac{b^2}{r_S^2}}}{r_S^2}\nn\\
&&+\frac{3b^4\sqrt{1-\frac{b^2}{r_R^2}}}{r_R^4}+\frac{3b^4\sqrt{1-\frac{b^2}{r_S^2}}}{r_S^4},
\end{eqnarray}
and the infinite-distance deflection is
\begin{eqnarray}
\label{Istring3}
\alpha_{a_0}|_{\infty}&=&-\frac{a_0\pi\left(8+v^2\right)}{256b^8v^2}.
\end{eqnarray}

\begin{figure}[htb]
\begin{center}
\includegraphics[width=0.4\textwidth]{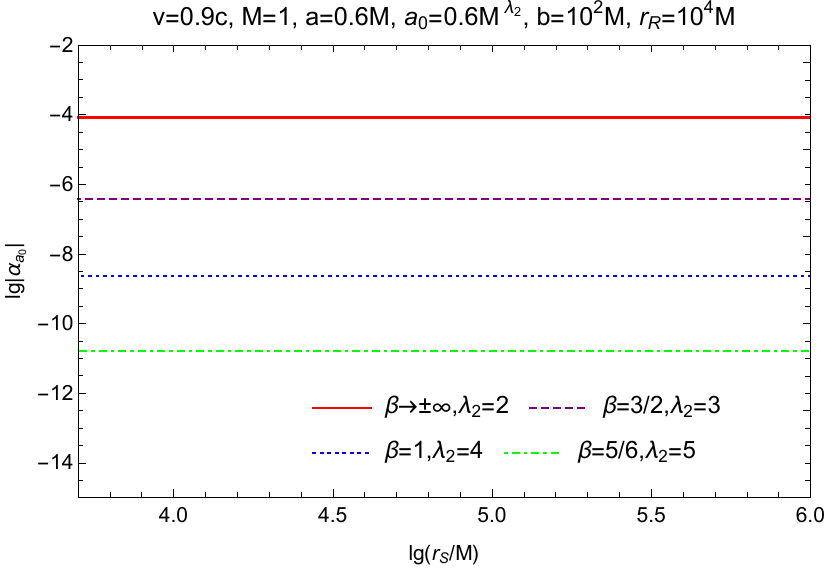}\hspace{0.04\textwidth}\\
\hspace{0.3\textwidth}\\
\caption{The deviation of finite-distance deflection angle from Kerr spacetime. The vertical axis takes the value $\lg |{\alpha_{a_0}}|$, and the horizontal axis takes $\lg(r_S/M)$, with $v=0.9c$, $M=1$, $a=0.6M$, $a_0=0.6M^{\lambda_2}$, $b=10^2M$, and $r_R=10^4M$. The red solid line, purple dashed line,  blue dotted line, green dash-dotted line correspond to $\lambda_2=2$, $\lambda_2=3$, $\lambda_2=4$, and $\lambda_2=5$, respectively.} \label{deviationangle}
\end{center}
\end{figure}

$\alpha_{a_0}$ versus source distance $r_S$ is plotted in Fig.~\ref{deviationangle}, where the vertical axis takes the value $\lg |{\alpha_{a_0}}|$, and the horizontal axis takes $\lg(r_S/M)$. We have set $v=0.9c$, $M=1$, $a=0.6M$, $a_0=0.6M^{\lambda_2}$, $b=10^2M$, and $r_R=10^4M$. The numerical results show that the effect of a cloud of strings decreases as $\lambda_2(\geq2)$ increases. In addition, it is shown that the impact of source distance $r_S$ on $\alpha_{a_0}$ is very small.

In the above, we did not consider the case of $\lambda_2=0$ (i.e. $\beta=0$), which describe the solution of rotating BH surrounded by a cloud of strings in GR. From~Eqs.~\eqref{KKerrGausss} or~\eqref{KerrGauss}, the term containing $a_0$ in Gaussian curvature disappears. Thus, it seems that the first order term of string parameter $a_0$ does not contribute to the deflection angle. However, from Eq.~\eqref{KKKerrangle}, one can see that the terms containing $a_0$ do not disappear when $\lambda_2=0$, and one can find that the result becomes
\begin{eqnarray}
\alpha_{a_0}&=&\frac{a_0}{2}\left(\arcsin \sqrt{1-\frac{b^2}{r_R^2}}+\arcsin \sqrt{1-\frac{b^2}{r_R^2}}\right).
\end{eqnarray}
For the infinite-distance case
\begin{eqnarray}
\alpha_{a_0}\mid_\infty&=&\frac{a_0}{2}\pi,
\end{eqnarray}
and it can be seen that this term is independent of the particle velocity $v$. In this assumption, the total infinite-distance deflection angle of massive particle reads
\begin{eqnarray}
\alpha|_{\infty}&=&\frac{2M\left(1+v^2\right)}{bv^2}+\frac{3\pi \left(4+v^2\right)M^2}{4b^2 v^2}\pm\frac{4M a}{b^2v}+\frac{a_0}{2}\pi+\mathcal O(\varepsilon^3).\nn\\
\end{eqnarray}
Let $v=1$, one can obtain the deflection angle of light in Kerr spacetime surrounded by a cloud of strings in GR, as follows
\begin{eqnarray}
\label{lightyuyuyu}
\alpha|_{\infty}&=&
\frac{4M}{b}+\frac{a_0}{2}\pi\pm\frac{4M a}{b^2}+\frac{15\pi M^2}{4b^2}+\mathcal O(\varepsilon^3).
\end{eqnarray}
In this equation, the first three items are consistent with Eq.~(42) in Ref.~\cite{Jusufi:monopole}.

\section{conclusion} \label{conclusion}

In this paper, the Kerr solution surrounded by a cloud of strings in Rastall gravity model has been obtained by using NJA without complexification. The influence of string parameter $a_0$ on BH's thermodynamic properties has been discussed. As a main result of this work, the gravitational lensing of massive particles for the Kerr solution obtained here is further considered by applying a geometric and topological method. For the case that the receiver and the source are infinitely far from the lens object, the GB theorem is applied to an osculating Riemannian space in which the particle ray is a geodesic, and the result related to string parameter $a_0$ is described by Gamma function in Eq.~\eqref{infangle}. For the case that the receiver and the source are finitely far from the lens object, the GB theorem is applied to a generalized Jacobi metric space. The particle ray is not a geodesic and its geodesic curvature should be considered to study the gravitational deflection. The result related to string parameter $a_0$ is described by hypergeometric function in Eq.~\eqref{KKKerrangle}. It should be noted that the finite-distance deflection angle in Eq.~\eqref{KKKerrangle} can lead to the infinite-distance angle~\eqref{infangle}, which shows that the same result for the infinite-distance deflection angle of massive particles can be obtained with the two methods. On the other hand, we find that the term containing $a_0$ in Gaussian curvature vanish when $\beta=0$ ($\lambda_2=0$). However, $\beta=0$ does not lead to this term vanishing in result. Furthermore, considering light case and setting $\lambda_2=0$, our result agrees with the result by rotating BH surrounded by a cloud of strings in GR.

\acknowledgements
The authors are grateful to the anonymous referees for their insightful comments and suggestions. Z. L. would like to thank Prof. Junji Jia for useful discussions, and Yaoguang Wang, Haotian Liu and Likang Fu for kind help, in Wuhan University. This work was supported by the National Natural Science Foundation of China (Grants No.~12047576 and No.~11947404), and the Fundamental Research Funds for the Central Universities (Grant No. 2682021ZTPY050).

\end{document}